# Interfacing a quantum dot spin with a photonic circuit

I.J. Luxmoore[1], N.A. Wasley[1], A.J. Ramsay[1,2], A.C.T. Thijssen[3], R. Oulton[3], M. Hugues[4], S. Kasture[5], Achanta V. G.[5], A.M. Fox[1] and M.S. Skolnick[1]

[1]*Department of Physics and Astronomy, University of Sheffield, Sheffield, S3 7RH, UK*

[2]*Hitachi Cambridge Laboratory, Hitachi Europe Ltd, Cambridge CB3 OHE, UK*

[3]*H.H. Wills Physics Laboratory, Tyndall Avenue, Bristol BS8 1TL, UK*

[4]*Department of Electronic and Electrical Engineering, University of Sheffield, Sheffield, S1 3JD, UK*

[5] *DCMP & MS, Tata Institute of Fundamental Research, Mumbai 400 005, India*

**A scalable optical quantum information processor[1-3] is likely to be a waveguide circuit[4] with integrated sources[5-7], detectors[8], and either deterministic quantum-logic or quantum memory elements[2,9]. With microsecond coherence times[10,11], ultrafast coherent control[12,13], and lifetime-limited transitions[14], semiconductor quantum-dot spins are a natural choice for the static qubits. However their integration with flying photonic qubits requires an on-chip spin-photon interface, which presents a fundamental problem: the spin-state is measured and controlled via circularly-polarised photons[12,13], but waveguides support only linear polarisation. We demonstrate here a solution based on two orthogonal photonic nanowires, in which the spin-state is mapped to a path-encoded photon, thus providing a blue-print for a scalable spin-photon network[15]. Furthermore, for some devices we observe that the circular polarisation state is directly mapped to orthogonal nanowires. This result, which is physically surprising for a non-chiral structure, is shown to be related to the nano-positioning of the quantum-dot with respect to the photonic circuit.**

Photons are the most robust carriers of quantum information and the most easily manipulated at the single qubit level[1]. With the recent implementation of Shor's Algorithm using several one- and two-qubit gates in a single waveguide circuit[3], linear optical quantum computing is an early front runner. In such a linear optics approach, measurements are used to implement two-photon gates resulting in a probabilistic operation[2]. However, without a deterministic single photon nonlinearity[16] or a quantum memory, such an approach is intrinsically un-scalable[2,9]. This can be overcome with a solid-state emitter, for example quantum dots (QDs), nitrogen-vacancy centres in diamond or impurity centres in semiconductors, which can be integrated within the waveguide circuit.

III-V semiconductor QDs are promising solid-state quantum emitters, with strong optical dipole and lifetime limited radiative recombination[14]. Highly tuneable QDs[17,18] are easily integrated with photonic structures: non-classical light sources[5-7], strong-coupling[19,20], indistinguishable photons from two remote QDs[21,22] and on-chip integration with single photon detectors[8] have all been demonstrated. The QD states most suitable as storage qubits are the spin eigenstates of an electron or hole, with intrinsic coherence times in the microsecond regime[10,11]. Optical detection and manipulation[12,13] of the spin is achieved by mapping to the circular polarisation state of a photon, i.e. $(\alpha|\uparrow\rangle+\beta e^{i\phi}|\downarrow\rangle) \Rightarrow (\alpha|\sigma^-\rangle+\beta e^{i\phi}|\sigma^+\rangle)$. However, interfacing to the spin ($\uparrow,\downarrow$) is problematic in a planar waveguide structure, where only the x or y component of the left ($\sigma^-$) and right ($\sigma^+$) circularly polarised light will propagate, severely constraining the scalability of a quantum network which exploits QD spins as static qubits.

In this work we employ a crossed photonic nanowire waveguide device, where the polarisation of a photon emitted by a QD at the intersection is converted to a path-encoded state, with the x(y)-polarisation component transmitted along the y(x)-direction waveguide. By recombining these waveguides the polarisation of the photon can be recovered and the spin-state of the QD deduced. Fig. 1(a) shows a scanning electron microscope image of a prototype spin-photon interface. This device consists of two orthogonal free-standing waveguides with a width of ~200nm connected to four out-couplers[23]. The waveguides are fabricated from a 140nm thick GaAs layer containing a single layer of InGaAs QDs at its centre (see Methods for details). A QD located at the centre of the waveguide intersection will coherently emit the x(y)-polarisation component of a circularly-polarised state into the waveguides aligned along the y(x)-directions respectively. By collecting both polarisation components, whilst retaining their relative phase, the full polarisation state of the photon is mapped to a path-encoded state. On recombining the light from the waveguides, the polarisation state of the photon can be reconstructed at another point in the plane, hence enabling on-chip transfer of spin information.

Finite difference time domain simulations (FDTD) simulations reveal that the spin to path conversion is sensitive to the QD position. We present results from two devices. For device-A, consistent with a QD located near the centre of the intersection, the coherent transfer of the full polarisation of the photon emitted by the QD to the path-encoded state is demonstrated. By contrast, in device-B, consistent with the QD located off-centre, the $\sigma^+$ and $\sigma^-$ polarisation components are mapped to different waveguides for direct read-out of the up/down state of the QD spin. Unexpectedly, in this case inversion symmetry between propagation along forward and backward aligned waveguides is broken. This result is in good agreement with the FDTD simulations, which show that an arbitrary spin state can still be transferred via a pair of waveguides when the QD is off-centre making the technique robust to QD alignment accuracy.

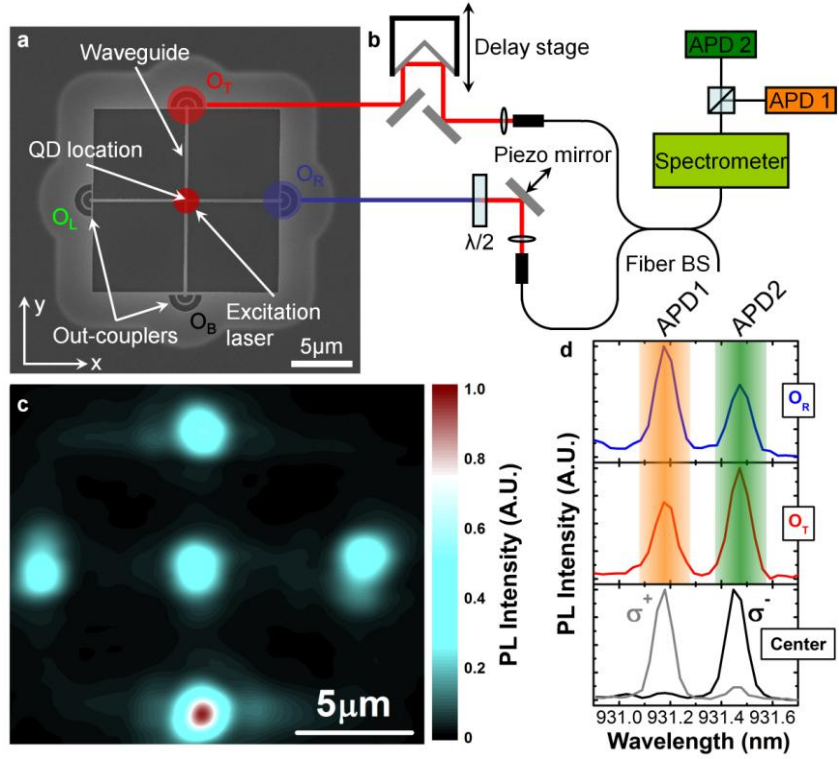

**Figure 1** Prototype spin to guided-photon interface and experimental set-up. (a) Scanning electron microscope image of the device. Two orthogonal ridge-waveguides are excited at their intersection by a CW laser, linearly polarised at 45° to the x-axis. QD emission into the two waveguides is measured via out-couplers, which scatter light into the z-direction. $O_T$, $O_R$, $O_B$ and $O_L$ signify the top, right, bottom and left out-couplers respectively. (b) Schematic diagram of the optical set-up used in the spectroscopy and interference experiments. (c) PL intensity map, integrated over the QD distribution, recorded from the spin-photon interface, by scanning the collection fiber with the excitation laser fixed at the intersection. (d) PL spectra recorded from a single Zeeman split QD line at the intersection and two out-couplers of device-A. At the intersection the transitions of the Zeeman doublet are found to be right and left circularly polarised, as expected. From the orthogonal out-couplers, $O_R$ and $O_T$, both $\sigma^+$ and $\sigma^-$ transitions are observed.

The demonstration of the principle proceeds by several steps. First, a laser is used to excite the QD wetting layer at the intersection. A map of the photoluminescence (PL), integrated over the QD ensemble, is presented in Fig. 1(c). Strong emission is observed from all four out-couplers and the intersection. This verifies that emission from the QDs excited at the waveguide intersection is transmitted along the waveguides and scattered vertically by the out-couplers.

The next step is to identify a device in which a single QD emits into two orthogonal waveguides. The structure is again excited at the intersection and PL spectra measured at the intersection and the out-couplers, labelled $O_R$ and $O_T$ in Fig. 1(a), are compared. A magnetic field, B=3T is applied normal to the sample plane, so that the $\sigma^+$ and $\sigma^-$ polarised transitions can be identified by their characteristic energies. Fig. 1(d) shows the µPL spectrum for the Zeeman split doublet originating from a QD located at the intersection of device-A. When observed vertically from the intersection, polarisation sensitive detection confirms that the two transitions are right and left circularly polarised. Both transitions can also be observed from all four out-couplers, with the spectra recorded from $O_R$ and $O_T$ shown in Fig 1(d). The contrast $C_\alpha$ between the $\sigma^\pm$ lines observed from the out-couplers, defined as $C_\alpha = (I^+ - I^-)/(I^+ + I^-)$, where α=T,R, are $C_R$=-0.19, and $C_T$=+0.19 respectively. This is consistent with mapping horizontal and vertical linear polarisations to the orthogonal waveguides, and is close to the desired value of $C_\alpha$=0. To verify that the QD is a single photon emitter, the second-order correlation function is measured and clear anti-bunching observed (see Supplementary Information).

Finally, to demonstrate that the full polarisation of the emitted photon is coherently mapped to the path-encoded state, we show that the light emitted from two orthogonal out-couplers is mutually coherent. To perform the interference experiment, two fibers are positioned to collect emission from $O_R$ and $O_T$. The two paths are recombined at a fiber beamsplitter, as depicted in Fig. 1(b), forming an interferometer. The interference fringes are recorded simultaneously for the $\sigma^+$ and $\sigma^-$ transitions by varying the path difference by a few wavelengths. An example of this data, where the $\sigma^\pm$ signals are in anti-phase is presented in Fig. 2(a). To deduce the relative-phase, $\Delta\phi$ between the $\sigma^\pm$ lines for each coarse time-delay $t_d$, a phase-plot of $\Gamma^+$ against $\Gamma^-$ is constructed (Fig. 2(b)).

Fig. 2(d) displays phase-plots at five different time delays. As the delay is varied from -0.8ps to 4.2ps, the ellipticity of the plot evolves from a straight line with negative gradient ($\Delta\phi=\pi$) to a straight line with positive gradient ($\Delta\phi=0$), via a circle ($\Delta\phi=\pi/2$), as a result of the Zeeman splitting between the $\sigma^\pm$ lines. Figure 2(c) presents a plot of $\Delta\phi$ as a function of $t_d$. $\Delta\phi$ oscillates with a period of 10ps, corresponding to the Zeeman splitting of 0.41meV. To determine $t_d=0$, the interference contrast of the entire QD ensemble, effectively a broadband light source, is measured and plotted in Fig. 2(c). At zero time-delay, the relative phase between the $\sigma^+$ and $\sigma^-$ lines is deduced to be $0.91\pi$, close to the expected value of $\pi$, the difference in the relative phase between the x and y polarised components of $\sigma^\pm=(x\pm iy)$ light. We note that for a path delay of $\lambda/4$, the $\sigma^+$ and $\sigma^-$ lines will exit via opposite ports of the fiber beamsplitter, which can be implemented for in-plane read-out of the spin up/down state of the QD. The above experiments demonstrate that for device-A, the polarisation of the photon emitted by the QD, including the relative phase between the polarisation components, and hence the full spin-state of the QD, is coherently mapped to a path-encoded state carried by the orthogonal waveguides.

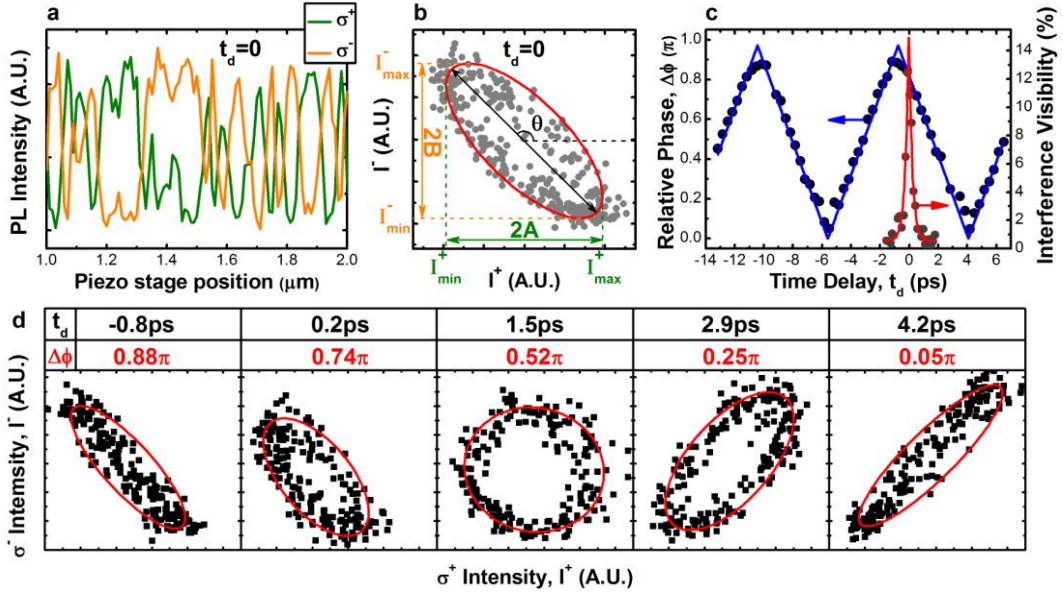

**Figure 2** Interference measurements of device-A. (a) Interference fringes, for $\sigma^+$ and $\sigma^-$ light detected from $O_R$ and $O_T$, recorded as the piezo mirror position is varied at $t_d=0$. (b) Intensity plot of the interference data presented in (a) showing that the $\sigma^+$ and $\sigma^-$ transitions are $\sim\pi$ out of phase at $t_d=0$ and can therefore be identified using an interference measurement in an in-plane architecture. The red line shows a fit of an ellipse to the data, which is used to extract the relative phase between the $\sigma^+$ and $\sigma^-$ transitions, from $\Delta\phi = cos^{-1}\left[\frac{A}{B}\tan(\theta)\right]$. (c) The left hand axis shows a plot of $\Delta\phi$ as a function of time delay. The blue line is a triangular waveform fit to the data. The right hand axis plots the visibility of the white light interference of the QD distribution, which is fitted with the function, $V(t) = V_0 + V_1 e^{(t_d-t_0)/\tau}$ to determine $t_d=0$. (d) Intensity plots of the interference fringes recorded simultaneously from the $\sigma^+$ and $\sigma^-$ transitions at different delay times of the interferometer. The solid red lines show elliptical fits to the data.

Device-B demonstrates markedly different behaviour with in-plane spin read-out without the need for an interferometer. Fig. 3(a) and (b) show PL spectra recorded for the Zeeman split doublet from the four out-couplers of device-B with B=4T. When measuring from the out-couplers, the $\sigma^+(\sigma^-)$ polarised light is only observed from $O_R$ and $O_B$ ($O_L$ and $O_T$). The contrasts extracted from Fig. 3(a) are $C_R=0.92$ and $C_L=-0.93$, which corresponds to the direct read-out of the spin state of the QD in an in-plane geometry.

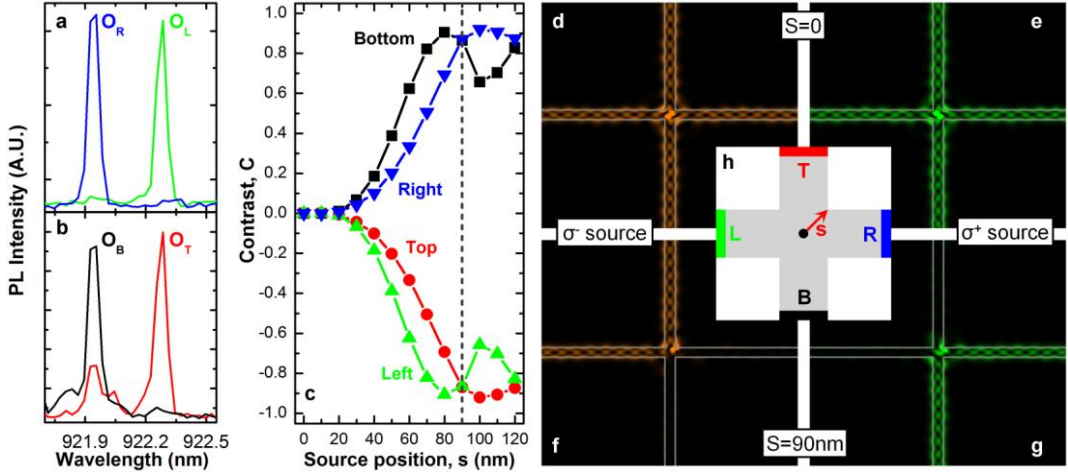

**Figure 3** Photoluminescence measurements for device-B and a numerical investigation into the effects of QD position on device operation. (a) PL spectra recorded from out-couplers $O_R$ and $O_L$ showing pronounced asymmetry between the two anti-parallel directions. (b) PL spectra recorded from $O_B$ and $O_T$, again showing the pronounced asymmetry of (a). (c) FDTD simulations showing the contrast of $\sigma^\pm$ polarised light in each waveguide for a source located at a distance, $s$ along the diagonal from the centre of the intersection. The dashed grey line indicates the location of the source for the simulations shown in (f) and (g), which reproduces well the emission properties of device-B. (d) - (g) Electric field intensity, $|E|^2$ at the centre of the waveguides for different source locations and polarisations 0.65ps after the CW source begins to emit. (d) $\sigma^-$ polarised source located at centre. (e) $\sigma^+$ polarised source located at centre. (f) $\sigma^-$ polarised source located off-centre at s=90nm. (g) $\sigma^+$ polarised source located off-centre at s=90nm. (h) Diagram illustrating the position of the source within the waveguide intersection as used in the simulations shown in (c).

To understand the different spin to path conversion properties of devices A and B, we use FDTD simulations to investigate the effect of QD position (see Methods). A circularly polarised dipole source is located in the intersection at a distance $s$ from the centre along the diagonal, as illustrated in Fig. 3(h), and the power transmitted into the four waveguides calculated. Fig. 3(c) plots the calculated contrast in the waveguides as a function of *s*, which is strongly dependent on source position. Close to the centre, light is coupled equally into all four waveguides, as observed for device-A. If however the QD is located at s≈90nm, the $\sigma^+$($\sigma^-$) polarised light is directed along the bottom and right (top and left) waveguides, reproducing the experimentally observed behaviour of device-B. From the simulations, we infer that for device-A, s<~50nm, and for device-B, s≈90nm. The operation of the two devices is illustrated by the electric-field distribution for light emitted by $\sigma^\pm$ polarised dipoles as presented in Figures 3(d, e) and (f, g) for s=0 and s=90nm respectively. Further calculations show that, despite the different spin to guided photon maps resulting from QD location, it is possible to deduce any arbitrary spin state via a measurement of the intensity and phase at two output ports. This implies that following characterization of the device, spin to guided-photon operation can be achieved for a range of QD positions (see Supplementary Information).

In device-B the observations that a non-chiral photonic structure separates the $\sigma^\pm$ components of the QD emission and breaks inversion symmetry between propagation in opposite directions are at first sight surprising. However, the phenomenological behaviour of this device can be explained as an interference of the light emitted by an x and y-polarised dipole, using coupled mode theory of a waveguide junction where the intersection is treated as a low-Q cavity[24], to deduce the relative phase of light emitted into each waveguide (see Supplementary Information). A key conclusion of this work is that the way in which a spin optically couples to a photonic device is sensitive to the QD position and is therefore an important design consideration for any spin-photon interface.

In summary, we have presented a scheme for interfacing an optically addressed spin qubit to a path-encoded photon using a crossed waveguide device. We have demonstrated the operation of this device in two regimes dependent on the location of the QD and shown that it can be used for in-plane transfer and read-out of spin information. Future directions include the use of nanocavities to enhance the light-matter interactions[25] and on-chip read-out using integrated single photon detectors[8]. This work provides a blue-print for the construction of a scalable on-chip network of solid-state spins; the next step is a demonstration of the remote entanglement[26] of two on-chip spins, which can be realized using two spin-photon interfaces and an integrated optical circuit.

## Methods

**Sample Growth and Fabrication**

The samples used in this study are grown by molecular beam epitaxy (MBE) on undoped GaAs (100) wafers. The wafer consists of a 140nm GaAs waveguide slab with a single layer of nominally InAs QDs at its centre, separated from the GaAs buffer layer by a 1μm thick $Al_{0.6}Ga_{0.4}As$ sacrificial layer. A rotation stop is included during the InAs deposition in order to achieve a low QD density. This results in a grading of the QD density across the sample and a minimum measured density of ~$1x10^9 cm^{-2}$. This area of the sample is then employed to fabricate the devices investigated here. Electron beam lithography is used to define the waveguide devices and the GaAs slab layer etched using a chlorine based reactive ion etch (RIE). Finally, hydrofluoric acid is used to selectively remove the $Al_{0.6}Ga_{0.4}As$ layer from beneath the waveguides leaving the free-standing waveguide structure.

**Optical Measurements**

The optical measurements, shown schematically in Fig. 1(b), are performed in a low temperature confocal microscope system equipped with a superconducting magnet. The photoluminescence is excited using an 855nm continuous wave (CW) diode laser, linearly polarised at 45° to the x-axis, as defined in Fig. 1(a) and focused to a ~1μm diameter. The spatial selection is achieved with a pair of single mode optical fibers which can be independently positioned in the image-plane to simultaneously collect emission from two discrete locations on the sample. For the interference experiments the light from the two out-couplers is coupled into a single mode fiber beamsplitter. The light from each output port of the fiber beam-splitter is filtered with a 0.55m spectrometer and monitored using a pair of silicon avalanche photo-diodes (APDs), which can be individually tuned to detect two different wavelengths simultaneously. The path delay in the interferometer is modulated using a linear delay stage and piezo-actuated mirror.

**Finite Difference Time Domain Simulations**

Finite difference time domain simulations are performed using MEEP (MIT Electromagnetic Equation Propagation)[27]. The simulated device consists of two 140nm×200nm GaAs waveguides surrounded by air, as presented in Fig. 3(h) (not to scale). A dipole source of arbitrary polarisation is created using a superposition of two broadband current sources orientated at 90° and located within the waveguide intersection a distance *s* from the centre. The power coupled into each waveguide is calculated by integrating the Poynting vector over a plane at the exit of the waveguide, 4.2μm from the centre: $P(\omega) = Re\{\hat{\boldsymbol{n}} \cdot \int E(r,\omega)^* \times H(r,\omega) d^2 r\}$ where $\hat{\boldsymbol{n}}$ is a unit vector normal to the plane, *r* is a position in the plane, ω is the angular frequency and E and H are the Fourier transforms of the E and H-fields. In a similar way, the total power emitted by the dipoles is calculated by integrating the Poynting vector over six flux planes positioned at the faces of the computational cell. The ratios of the fluxes collected in the arms of the waveguides and at the faces of the computational cell give the coupling ratios, which are evaluated for a free-space wavelength of 922nm. With the source located at the centre of the intersection the coupling efficiency into each waveguide is ~14%, with losses of ~44%. Ideally the coupling into each waveguide would be close to 25%; improvements to increase the efficiency can be made via optimisation of the waveguide dimensions.

**Acknowledgements**

We thank EPSRC grant numbers EP/G001642 and EP/J007544 for support of this work and D. N. Krizhanovskii, L. R. Wilson, P. Kok and D. M. Whittaker for helpful discussions and comments on the manuscript.

## Supplementary Information

**1. Quantum Dot Properties**

To verify that the quantum dots (QDs) studied in this work are single photon emitters, we perform a photon correlation measurement using a Hanbury-Brown and Twiss (HBT) set-up. For device-A the QD emission is collected in the vertical direction from the intersection of the waveguides and filtered with a spectrometer, before being split with a beamsplitter and detected using a pair of APDs (see Fig. 1(b) of main article). The histogram recorded of the relative time delay, τ, between coincidence events on the two APDs is shown in Fig. S1(a) for the QD in device-A. The measurement shows anti-bunching with a multi-photon emission probability of $g^{(2)}(0)=0.28$ obtained with no background subtraction, clearly demonstrating the single-photon nature of the emission.

The interference measurements presented in Fig. 2 of the main article can be used to measure the coherence of the QD emission. The interference visibility at a given position of the delay stage is given by $V^+ = (I^+_{max} - I^+_{min})/(I^+_{max} + I^+_{min})$ and $V^- = (I^-_{max} - I^-_{min})/(I^-_{max} + I^-_{min})$ for the σ⁺ and σ⁻ transitions respectively (see Fig. 2(b) in main article). Figures S1(b) and (c) plot the decay of the interference visibility against the time delay for the σ⁻ and σ⁺ transitions, respectively. The data is fitted following ref. S1 and coherence times of 81ps and 68ps are extracted for the σ⁺ and σ⁻ transitions, respectively.

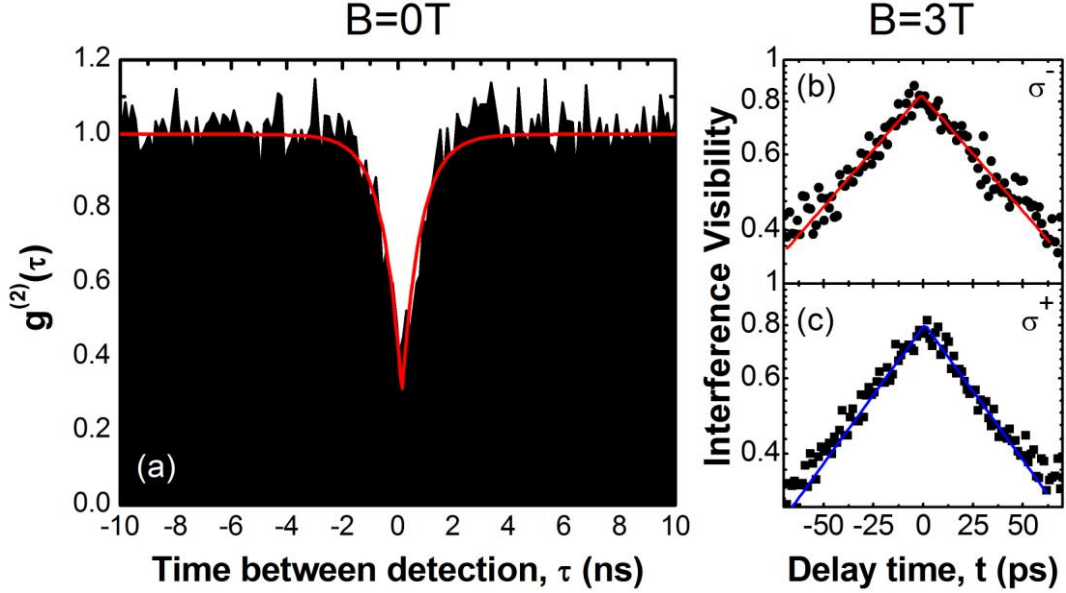

**Figure S1** (a) Photon correlation histogram of the QD in device-A recorded using an HBT set-up. The red line shows a fit to the data with the function $g^{(2)}(\tau) = \left(1 - (1-B)e^{-|\tau|/\tau_d}\right)$, where B=0.28 is the multi-photon emission probability and $\tau_d$=0.76ns is the lifetime of the QD. (b, c) Coherence time measurements of the $\sigma^-$ and $\sigma^+$ transitions, respectively. The solid red and blue lines show fits to the data following ref. S1.

Figure S2 presents the second-order *cross-correlation* between light emitted from output couplers $O_T$ and $O_R$ for device-B. The QD emission is filtered using a narrowband interference filter (FWHM of ~1.5nm). To correct for the large background signal resulting from the relatively large spectral window we use the equation, $g^2(\tau) = [C_N(\tau) - (1-\rho^2)]/\rho^2$, where $C_N(\tau)$ is the normalized data and $\rho$ is the signal to background ratio [S2]. The cross-correlation histogram for the QD studied in device-B is shown in Fig. S2 and shows anti-bunching with $g^{(2)}(0)$=0.04. This again demonstrates the single photon nature of the emission, but using photons collected by the waveguides.

The coherence time for the QD in device-B can be extracted as above from an analysis of the visibility decay curves. From fits to the data, coherence times of 100ps are found for both the σ⁺ and σ⁻ transitions, respectively. The measured coherence times of the two QDs are typical for InGaAs QDs in photonic structures [S3, S4].

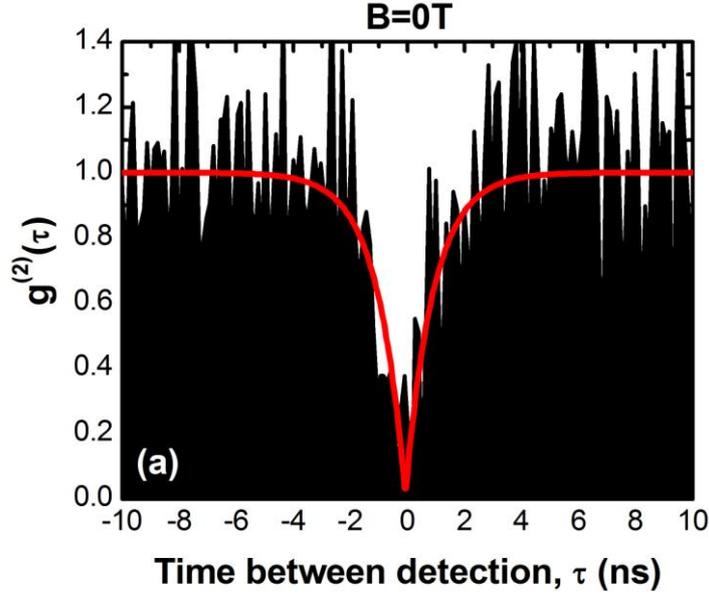

**Figure S2** Cross-correlation histogram recorded from $O_T$ and $O_R$ on device-B. The red-line shows a fit to the data of the function $g^{(2)}(\tau) = \left(1 - (1-B)e^{-|\tau|/\tau_d}\right)$, with B=0.04 and $\tau_d$=1.01ns.

## 2. Effect of QD position on the fidelity of the spin to path conversion

In this section we use FDTD simulations to verify that a one to one map between the polarisation of a photon emitted by a QD at the centre of the structure (device-A) and the path-encoded state is achieved. Furthermore, we ask to what extent device-B also performs a coherent map between polarisation and path state, which, following characterization of the device, can be corrected using a single qubit rotation.

To this end, the waveform of the $E_{x(y)}$ field in each waveguide, sampled at a point in the centre of the waveguide, equidistant from the centre of the intersection, is calculated for a QD at the centre, and an off-centre QD at s=90nm along the diagonal. An amplitude, $a$ and phase, $\varphi$ are found by fitting the waveform to a cosine, and these are used to construct a phasor, $ae^{i\varphi}$. We then choose two of the waveguides {T,R,B,L}, as defined in Fig. 3(h), to encode the photon state, $W$ for input polarisation, $p$ as, $W(p) = \{a_1 e^{i\varphi_1}, a_2 e^{i\varphi_2}\}$.

If the device performs a one-to-one map between polarisation and the path-encoded state, then orthogonal input polarisations will map to orthogonal path states. To quantify this as a measure of the quality of the state conversion, the dot product, $O(WG:position) = W(p).W(p')^*$, of the path-encoded states arising from orthogonal input polarisations, $p$ and $p'$ is calculated. This is then normalised to half the total intensity out of all four waveguides, with a value of zero ideal. To quantify the extent to which the spin-photon interface polarises the path-encoded state, we also calculate the total intensity of the light collected by the selected waveguide pair $I(WG:position)$, to check that this is independent of the input polarisation. The results are summarised in table S1. We use polarisation bases linear horizontal/vertical (H/V), linear diagonal/anti-diagonal (D/A), and right/left circular (σ⁺/σ⁻). The results presented in column 1 of table S1, confirm that the spin to path conversion is near perfect for a QD at the centre of the intersection. For a QD at s=90 nm from the centre along the diagonal, the H/V polarisations couple to all four waveguides, D/A couple to the TR/BL waveguides respectively, whilst σ⁺/σ⁻ couple to RB/LT waveguides respectively. This reflects the symmetries of the combined QD-photonic system, where the QD is considered as a dipole. To map the spin to a path-encoded state, we choose the TB waveguide-pair, since D and $\sigma^+$ (A and $\sigma^-$) polarisations map to the T (B) waveguide. As indicated in column 3 of table S1, for the off-centre case, it is also possible to map the polarisation to a path-encoded state with high fidelity. This suggests that whilst the state conversion is sensitive to the QD position, the unitary nature of the map is robust. Therefore for an off-centre QD the spin-photon map also works with high fidelity, albeit with an additional single qubit rotation needed to correct the spin to path map.

| Polarisation basis | O (TR; centre) | I (TR; centre) | O (TB; off-C) | I (TB; off-C) |
| --- | --- | --- | --- | --- |
| H/V | 1.5% | 1.00/1.00 | 1.6% | 1.03/0.96 |
| D/A | 0.005% | 0.99/1.01 | 6.8% | 1.00/1.00 |
| $\sigma^+/\sigma^-$ | 1.5% | 1.00/1.00 | 7.0% | 1.00/1.00 |

**Table S1** Calculated measures of the quality of state conversion for the case of on and off-centre QDs. O (WG; position) is the orthogonality, where a value of zero is ideal. I (WG; position) is the total intensity, where a value of one is ideal. WG refers to the waveguide-pair used. The polarisation bases used are linear horizontal/vertical (H/V), linear diagonal/anti-diagonal (D/A) and $\sigma^+/\sigma^-$ circular.

### 3. Physical explanation of spin read-out behaviour of device-B

In this section we present a physical interpretation of the spin read-out behaviour of device-B. This behaviour, in which inversion symmetry is broken for light propagating in opposite directions, may appear counterintuitive at first sight but can be explained using temporal coupled mode theory of a waveguide junction, where the intersection is treated as a low-Q cavity [S5]. We proceed by analysing the relative phase, $\varphi$ of the $\mathcal{H}_z$ magnetic field in the four waveguides for light emitted by a linearly polarised dipole located in the waveguide intersection.

H-polarised light can couple forwards into the top waveguide with phase 0, and backwards into the bottom waveguide with relative phase π without a change in polarisation. The π phase-shift arises from the fact that an oscillating dipole emits a $\mathcal{H}_z$ field with a sin(θ) dependence [S6]. For a source located at the centre of the intersection there is no coupling of the H-polarised light to the right and left waveguides. However, the simulations show that as the source is moved away from the centre; the light is increasingly coupled into the right and left waveguides. This coupling can be analyzed using the phenomenological temporal coupled mode theory [S5]. In this model, the intersection is treated as a weak cavity that couples to the waveguides. We assume that the horizontal dipole, $\mathcal{H}_H$ feeds the cavity mode M, and then leaks to the right and left waveguides. This is modelled by:

$$\frac{dM(t)}{dt} = -i\omega_0 M(t) - \frac{M(t)}{\tau} + C_1 \mathcal{H}_H(t),$$

where $\omega_0$ is the cavity resonance, $\tau$ the cavity lifetime, and $C_1$, $C_2$ are coupling coefficients that depend on the QD-position and

$$\mathcal{H}_R(t) = C_2 M(t),$$

where $\mathcal{H}_R$ is the $\mathcal{H}_z$-field in the R waveguide. If $\mathcal{H}_H(t) = \mathcal{H}_{H0} \exp(-i\omega t)$, then

$$\mathcal{H}_R(t) = \frac{C_1 C_2 \mathcal{H}_{H0} \exp(-i\omega t)}{i(\omega_0 - \omega) + \frac{1}{\tau}} = M_0 \exp(-i(\omega t + \varphi)),$$

and if $(\omega_0 - \omega)\tau \gg 1$, $\mathcal{H}_R(t)$ is retarded in phase by $\varphi = \pi/2$, if the cavity resonance is of low frequency. Hence, there is a π/2 phase shift for light emitted by a H-polarised dipole and scattered into the right or left waveguide. Therefore, for light emitted by the H-polarised dipole the phase in the four waveguides is $\varphi_H = \{0; \pi/2; \pi; \pi/2\}$. Similarly, for a vertically polarised dipole the phase in the four waveguides is $\varphi_V = \{\pi/2; 0; \pi/2; \pi\}$ and the relative phase $\Delta\varphi = \varphi_H - \varphi_V = \{-\pi/2; +\pi/2; +\pi/2; -\pi/2\}$. In other words, the horizontal polarisation is capacitively coupled to the right and left waveguides and hence there is a π/2 phase-shift.

In the case of σ$^{\pm}$ polarised emission, the source is a superposition of H and ±iV dipoles, which results in relative phases of $\{\pi; 0; 0; \pi\}$ and $\{0; \pi; \pi; 0\}$ between the light emitted by the H and V dipoles respectively. This leads to constructive interference for Δφ=0 and results in the σ$^{+}$ emission being predominantly emitted into the right and bottom waveguides and the σ$^{-}$ into the top and left waveguides, as observed in Fig. 3 of the main article. Table S2 summarizes the expected phase of the $H_z$-field from this analysis for various different dipole sources.

|  | Top | Right | Bottom | Left |
|---|---|---|---|---|
| **H** | 0 | π/2 | π | π/2 |
| **V** | π/2 | 0 | π/2 | π |
| **Δφ** | - π/2 | + π/2 | +π/2 | - π/2 |
| **σ⁺** | π | 0 | 0 | π |
| **σ⁻** | 0 | π | π | 0 |

**Table S2** Expected phase of $H_z$-field observed in top, right, bottom and left waveguides as emitted by dipoles of different polarisations for a source located at s=90nm.

We support this phenomenological explanation of device-B with simulation results. In Fig. S3 the waveforms of the $H_z$-field, as sampled by a probe at the centre of waveguides $\{T; R; B; L\}$ are calculated using FDTD. Both the H and V polarised dipoles are coupled into all four waveguides when the QD is positioned at s=90nm, whereas for s=0, the H(V) polarisations are only coupled to the top and bottom (right and left) waveguides respectively. The phase φ of the $H_z$-field can be extracted from a cosine fit, $\cos(\omega t + \varphi)$, to the waveforms shown in Fig. S3. The relative phases between light emitted by H and V dipoles, are $\{-0.34\pi; 0.36\pi; 0.65\pi; -0.64\pi\}$ for the four waveguides respectively, close to the predicted values of $\{-\pi/2; +\pi/2; +\pi/2; -\pi/2\}$.

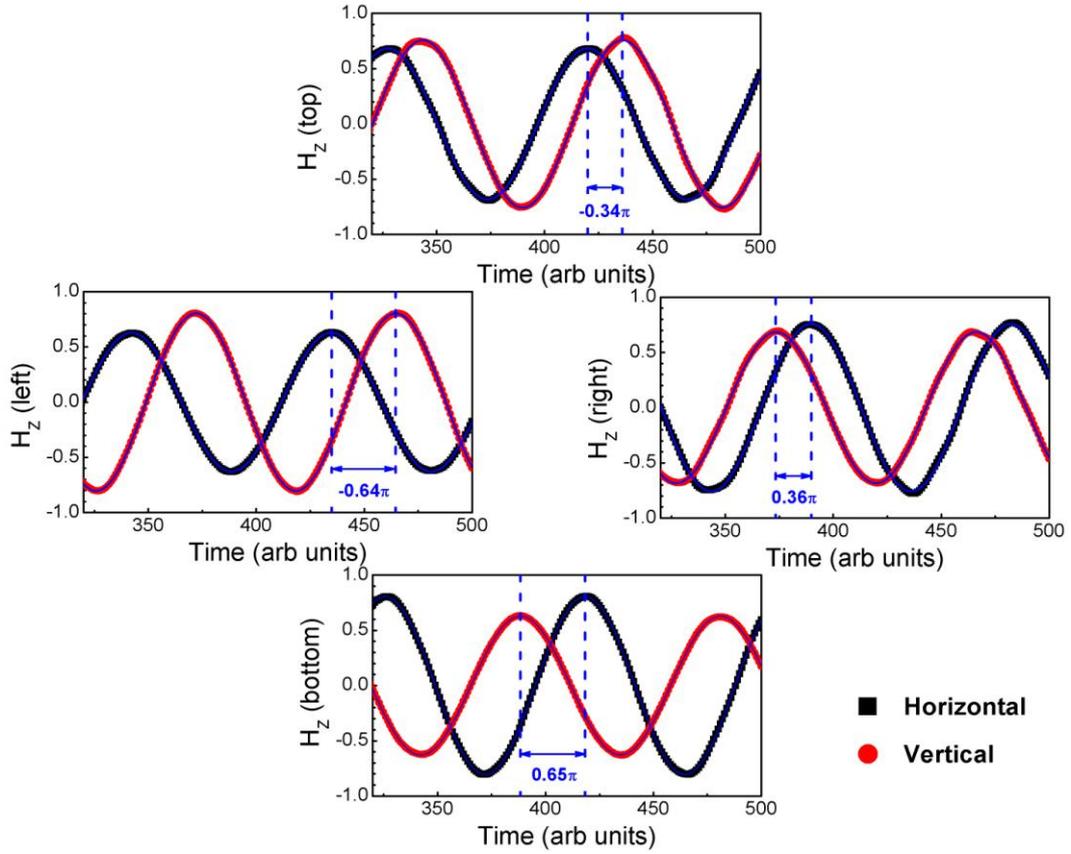

**Figure S3** Clockwise from top, calculated waveforms of the $H_z$-field components for the top, right, bottom, left waveguides, for H (black) and V (red) dipole orientations of a source located at s=90nm along the diagonal of the intersection. The relative phases of the waveforms $\Delta\varphi = \varphi_H - \varphi_V = \{-0.34\pi; +0.36\pi; +0.65\pi; -0.64\}$.

**References**

[S1] Berthelot, A. *et al.* Unconventional motional narrowing in the optical spectrum of a semiconductor quantum dot. *Nature Phys.* **2**, 759 (2006).

[S2] Brouri, R., Beveratos, A., Poizat, J.P. & Grangier, P. Photon antibunching in the fluorescence of individual color centers in diamond. *Opt. Letters* **25**, 1294 (2000).